# A Small Multi-Wire Telescope for High Energy Cosmic Ray Muon Detection

Maghrabi, Abdullrahnan, Al Enizy, Mohammed, Aldosari A, and Almuteri M

*National Centre For Applied Physics, King Abdulaziz City For Science and Technology, Riyadh 11442, Saudi Arabia,*
*amaghrabi@kacst.edu.sa*

## Abstract

Different types of ground-based detectors have been developed and deployed around the world to monitor and study CR variations. We have designed, constructed and operated a three layer small (20x20 $cm^2$) multiwire proportional chamber MWPC telescope for cosmic ray muon observations. In this paper, the technical aspects of this detector will be briefly discussed. The abilities of the telescope in detecting high nergy cosmic ray muons (primaries higher than 20 GeV) were established. The telescope performs well in this sense and showed comparable results with a 1 $m^2$ scintillator detector.

## 1. Introduction

Recently it has been established that the cosmic ray CR incident at the top of the atmosphere leads to changes in some atmospheric properties, which in turn affect global weather and climate[1-2]. These primaries are modulated over different time scales due to the solar and interplanetary process. Secondary cosmic ray particles observed at the sea level result from the interaction of the primary cosmic ray particles with atmospheric nuclei. The majority of these observed secondaries are muons due to their high survival probabilities[3].

A multiwire proportional chamber (MWC) is a type of gaseous detector known by its position resolution accuracy [4-5]. It is also recognized for its high counting rate. MWC detectors have several advantages in comparison with other types of gas detectors. For instance, a flexible detector shape can be designed with a high performance at a low cost [6].

The MWPC consists of an array of thin, parallel and equally spaced wires. Depending on the purpose of the measurements, the chamber is usually filled with an appropriate mixture of gases. The charged particles will ionize the gas atoms, producing pairs of electrons and ions. The produced ions and electrons will drift to their collecting electrodes, the cathode for the former and anode for the latter, producing an electric current relative to the energy of the incident particle [6-7].

As part of KACST radiation detector laboratory activities, we have constructed and operated a three layer small (20x20 $cm^{2)}$ MWPC detector. The prototype telescope succeeded in detecting high energy cosmic ray muons with acceptable accuracy. In this paper, the technical aspects of this system will be briefly discussed. Comparisons of the obtained results from this telescope with a 1 $m^2$ scintillator detector will be presented.

## 2. The detection system

The detection system, figure (1), consists of a set of MWPC, which is powered by a high voltage supplier. The outputs signals from the detector are pre-amplified using a custom-made preamplifier circuit. These signals are selected against the background noise using a discriminator unit. MT-10 tracking boards were used to determine the ionization position inside the detector. A Data Acquisition Card (MtRD Board) receives signals from both the preamplifier and discriminator chips and from the tracking boards and sends them to a RaspberryPi computer card.

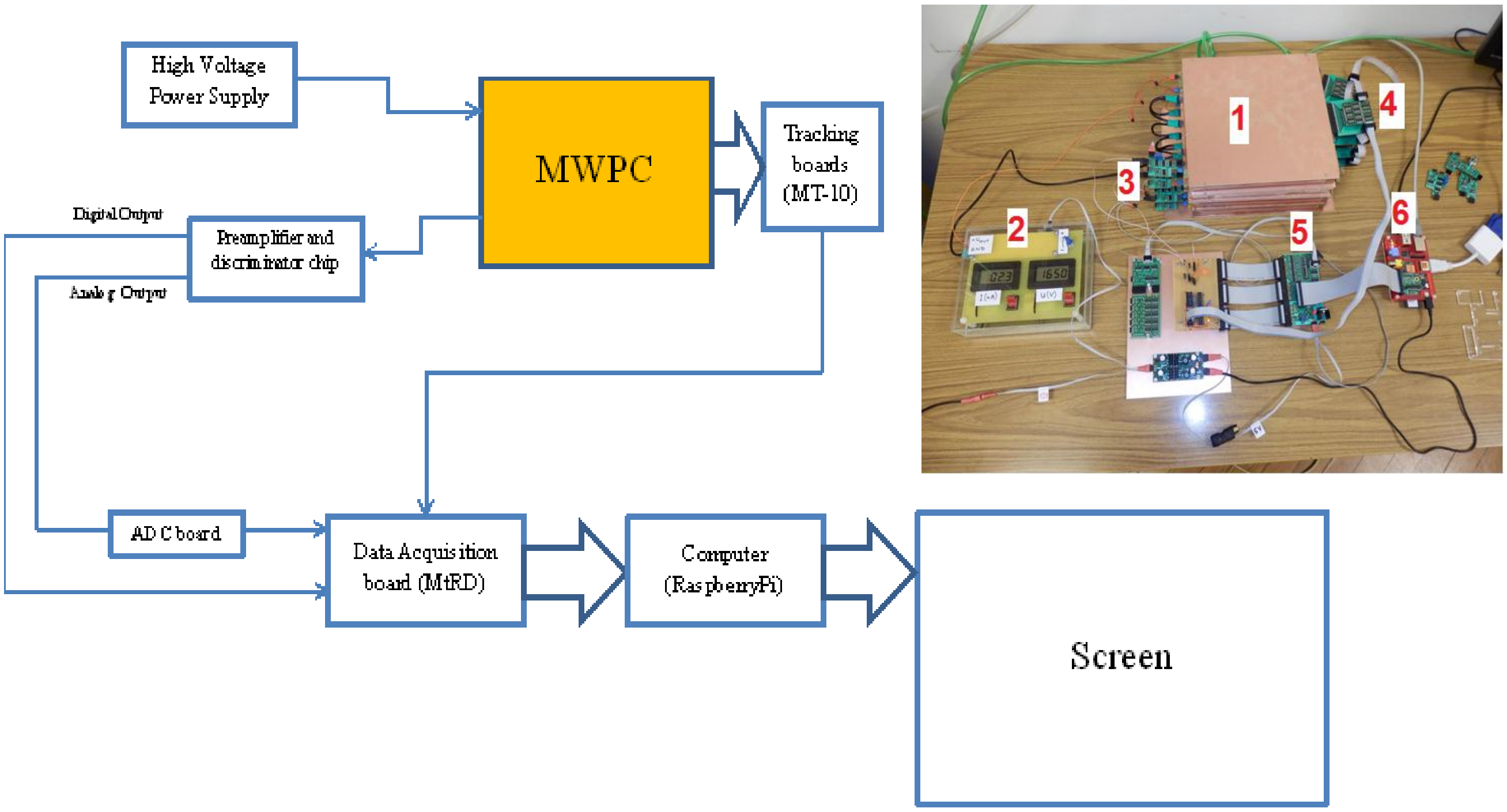

Figure 1: (a) schematic diagram of the developed MWPC (b) the components of the detector. 1: the MW champers; 2: HV power supply with 1500 to 1700 V; 3: preamplifier and discriminator chips; 4: tracking boards (or MT-10); 5: data acquisition board; and 6: RaspberryPi computer.

All the telescope components were arranged and constructed in the Physic Detector Group's laboratory at the National Center for Applied Physics (NCAP), King Abdulaziz City for Science and Technology (KACST), Riyadh, Saudi Arabia. The electronic components and the detector's materials were provided by Wigner institute, Hungary as part of a collaborative project in radiation detector constructions. Atmospheric pressure and laboratory temperatures were measured by a KACST weather station installed at the roof of the lab building [7].

The designed telescope is three layers of 20 × 20 MWPC stacked together. Each layer consists of an array of 16 anodes and 16 field wires. The sense wires (anodes), which are fed with HV, are made of 12 μm thick Tungsten whereas the field wires (ground potential) are made of 24 μm of copper. The anode wires are separated by 12 mm, and field wires are in between, as shown in figure (2a). The wires are placed on Plexiglas bars on both sides, stretched over the detector champers and fixed on the specified PCB spots (figure 2b). Due to their cost efficiency and safety reasons (a non-flammable, non-toxic mixture) a gas mixture of AR: $CO_2$ is used with a ratio of 80:20. The gas was left to flow through the detector for some time to adequately reduce the level of residual oxygen.

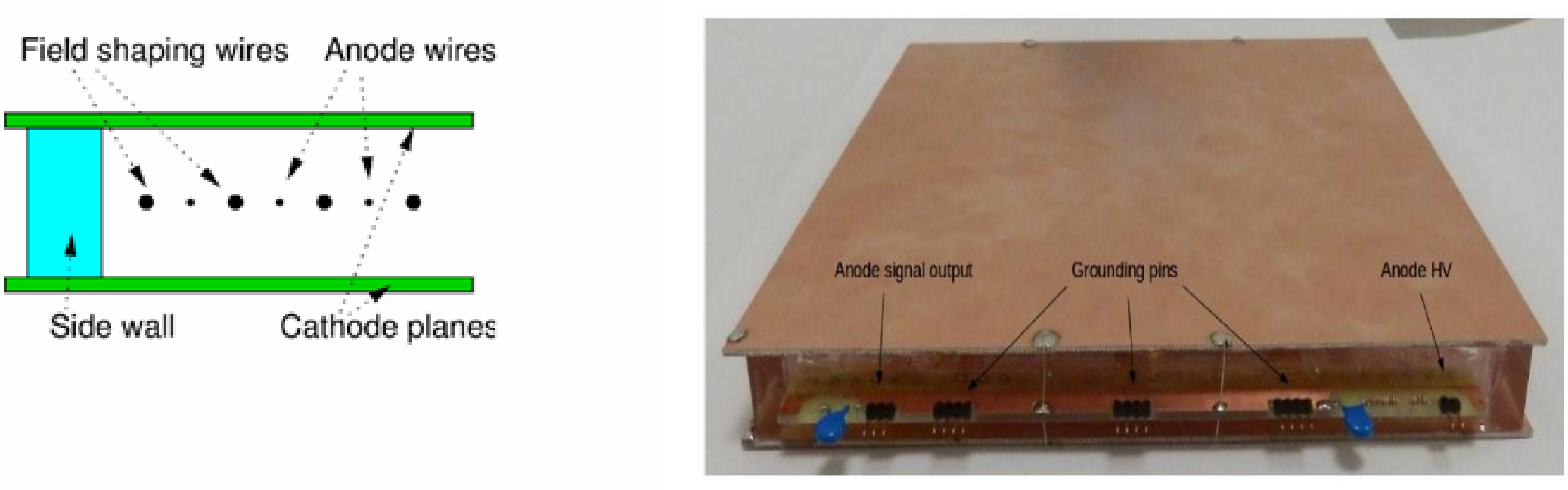

Figure 2: (a) schematic diagram shows the anode and field wires. (b) side view of a designed MWC.

The electronic components (figure 3) are then fixed to the detector. These are: the high voltage, preamplifier, discriminator, tracking boards, DAC and RaspberryPi computer card [8-9].

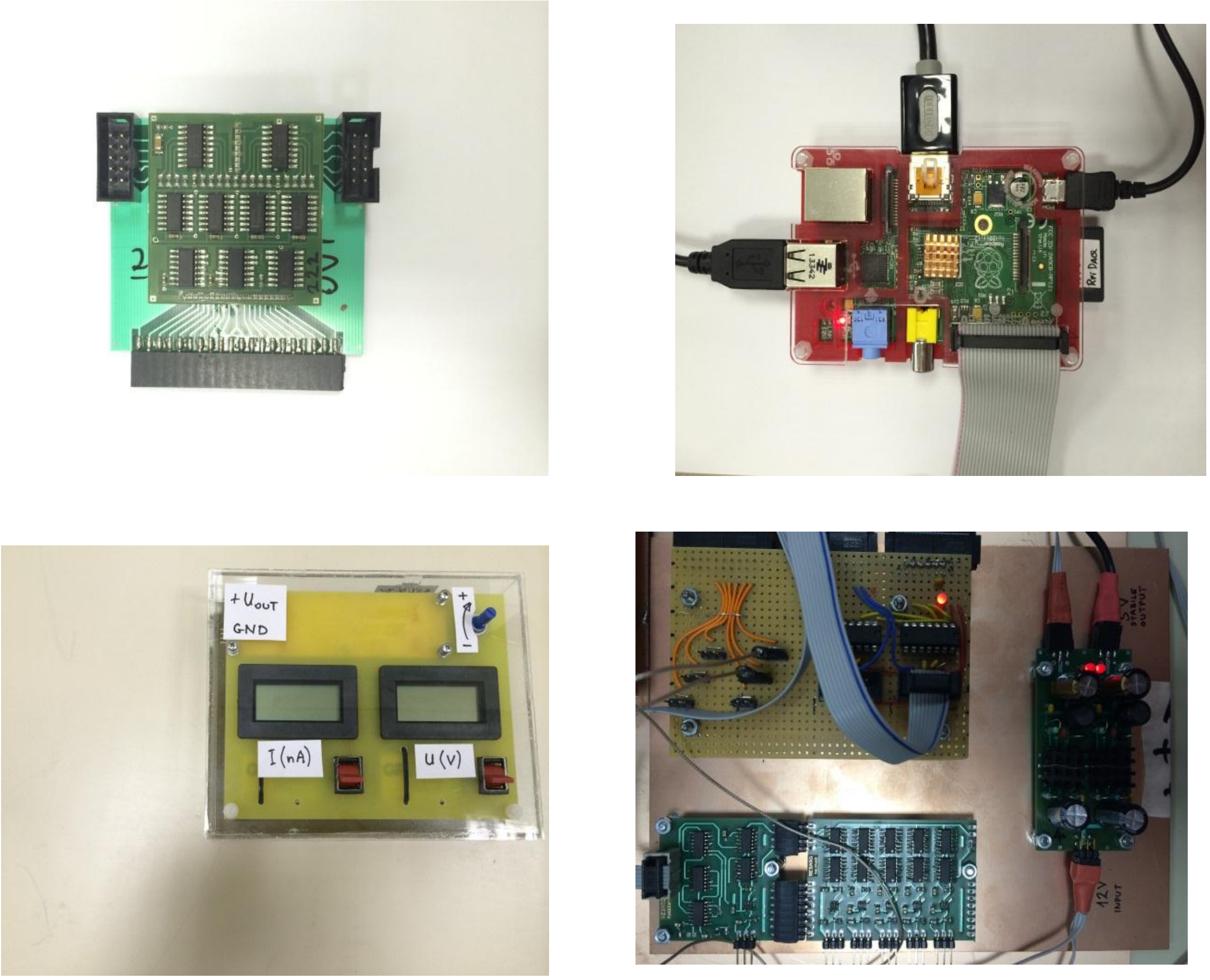

Figure (3) is an example shows some of the electronic parts of the detector. From top left: data acquisition board; RaspberryPi computer; : HV power supply ; tracking boards (or MT-10)

The detector performance can be characterized by its capability to distinguish the pulse height distribution caused by the cosmic ray muons from the distribution produced by the thermal electrons. This peak is called a Single Particle Peak (SPP), and its distribution is approximated by the Landau distribution.

The acquisition system allows us to specify the number of coincidences we are interested in. In our case, we set the trigger to record a particle that hits the three layers simultaneously (figure 4a). This means that we can remove the low energy background noise from the electrons and detect high energy muons with high accuracy. The coincidents from the sense wires are picked up if they are within a time window of 2 μs (pulse width observed in figure 4b and c). By setting up the discriminator circuit to a predefined threshold, CR muons will be detected if the signals from the three detectors are above this threshold, and the low-energy part of the spectrum that originates from the background radiation will be removed.

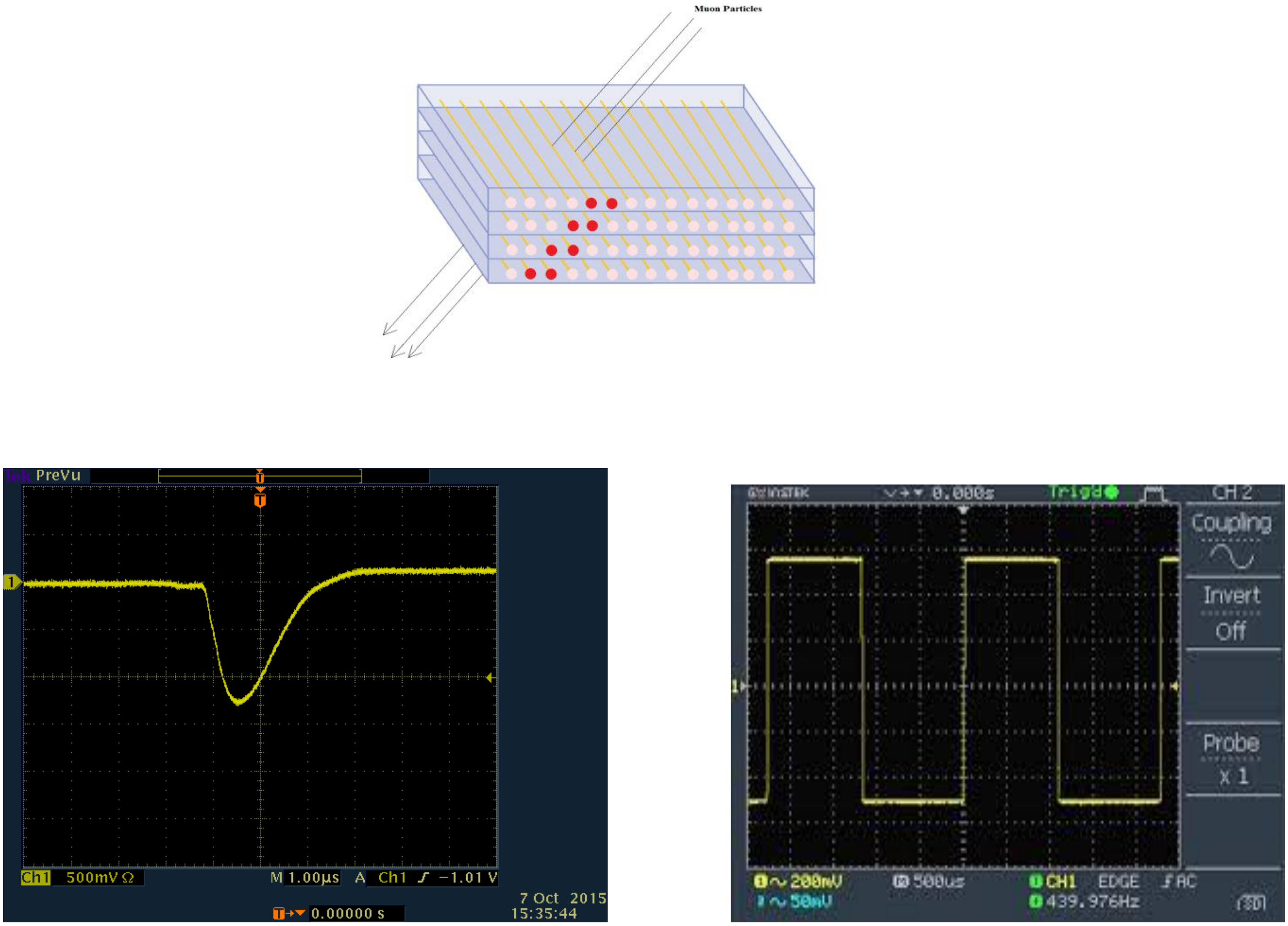

Figure 4: (a) schematic diagram of the principle of setting up the triggers. (b) analogue and (c) square output signals from the detector.

Figure 5 is a screen shot of the output from our detection system indicating the SPP corresponding to the cosmic ray muons as it separated from the background noise using the defined threshold.

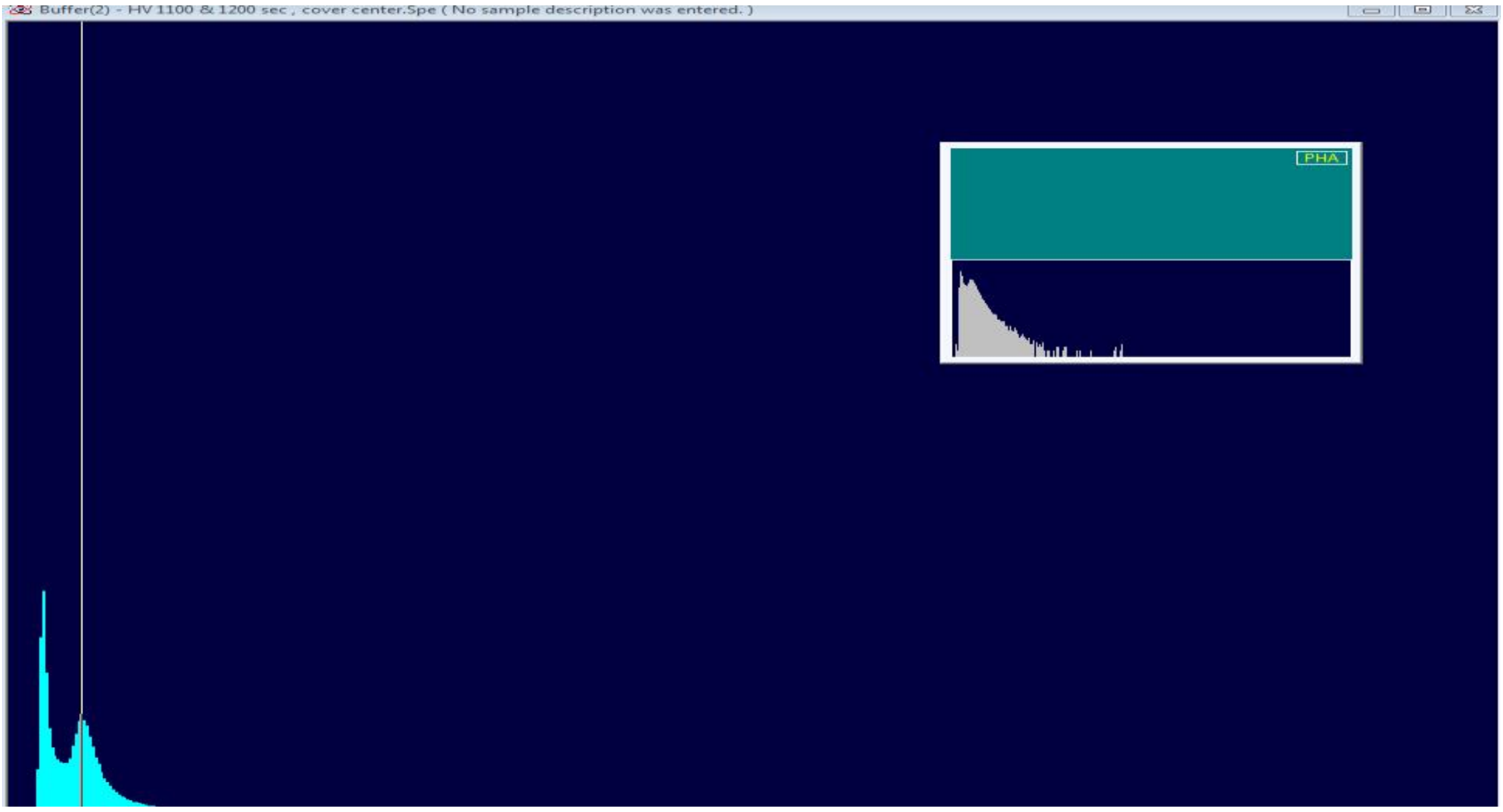

Figure 5: a muon peak separated from the background noise for an experiment of 10 hours.

# 3. Results

After the assembly of the detector and the completion of the necessarily tests, the detector was run for a period of two months starting in early September 2015. To avoid any heating effect of the detector's components, the system was operated at a stable lab temperature (23–25 $^{o}$C). The data were acquired every 1 ms and then averaged over one minute and over half an hour to reduce the statistical uncertainties and increase the accuracy of the counted muons. Figure (6) shows the rate of the muons detected by the developed MWC telescope for the period of 10 days. Data from our 1 m$^2$ scintillator detector are also included in the figure. It can be seen that, over the considered period of time, the behaviour of the recorded muons from the MWC is consistent with the behaviour of those collected by the larger scintillator detector most of the time.

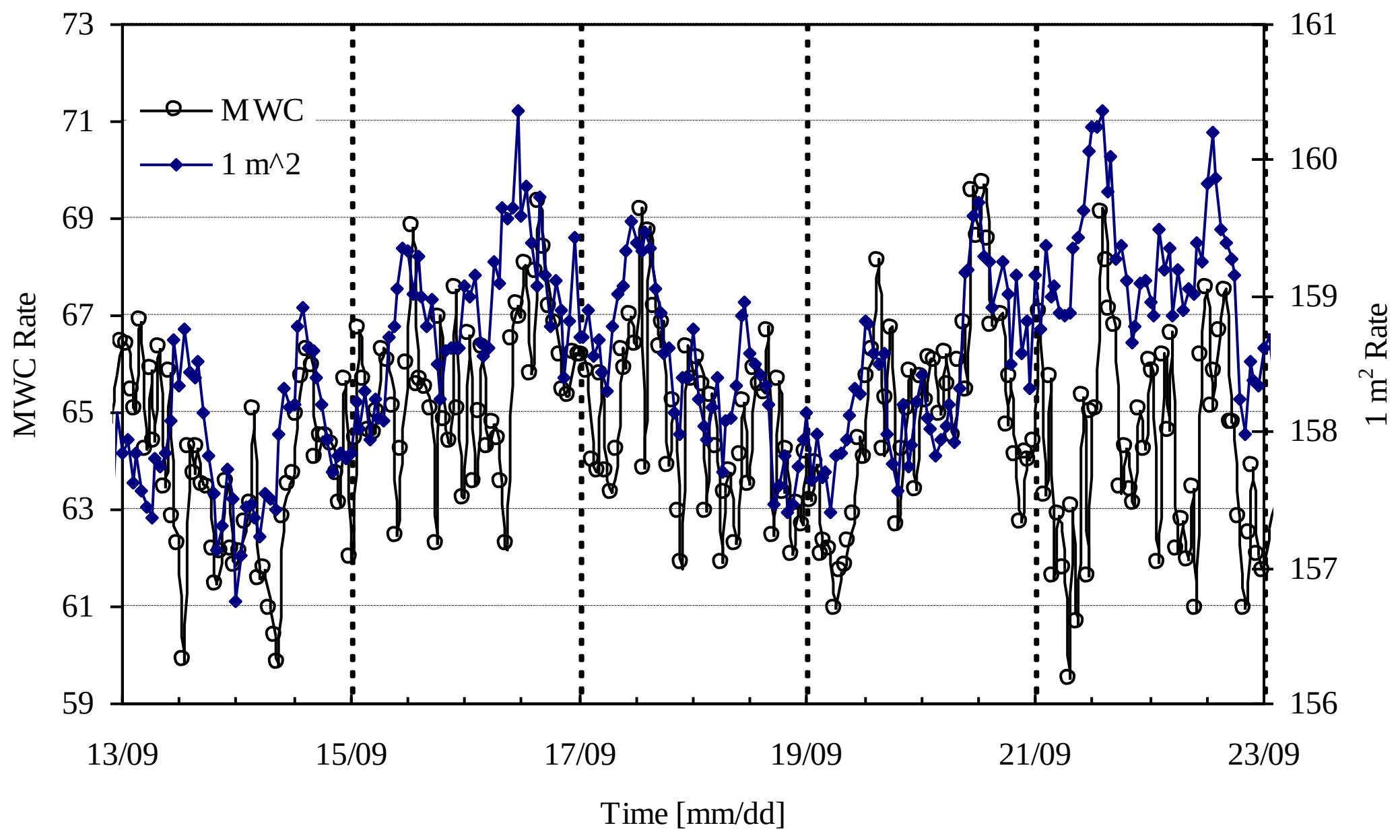

Figure 6: muons collected by the developed MWC between 13 and 23 September 2015 are compared with those observed by the KACST 1 $m^2$ scintillator detector.

It is well known that the rate of observed muons at the sea level is affected by some atmospheric variables, mainly atmospheric pressure and temperature. The atmospheric pressure represents the amount of the atmospheric materials that the muons should go through from their production level to the detector.  On the other hand, variations in the atmospheric temperature cause changes in its density, which affects the rate of the detected muons[10-11].

The relationship between the muon rate from MW telescope and the atmospheric pressure over a period of 25 days is presented in Figure (7). It can be seen that the muon count rate decreases as the pressure increases and vice versa.

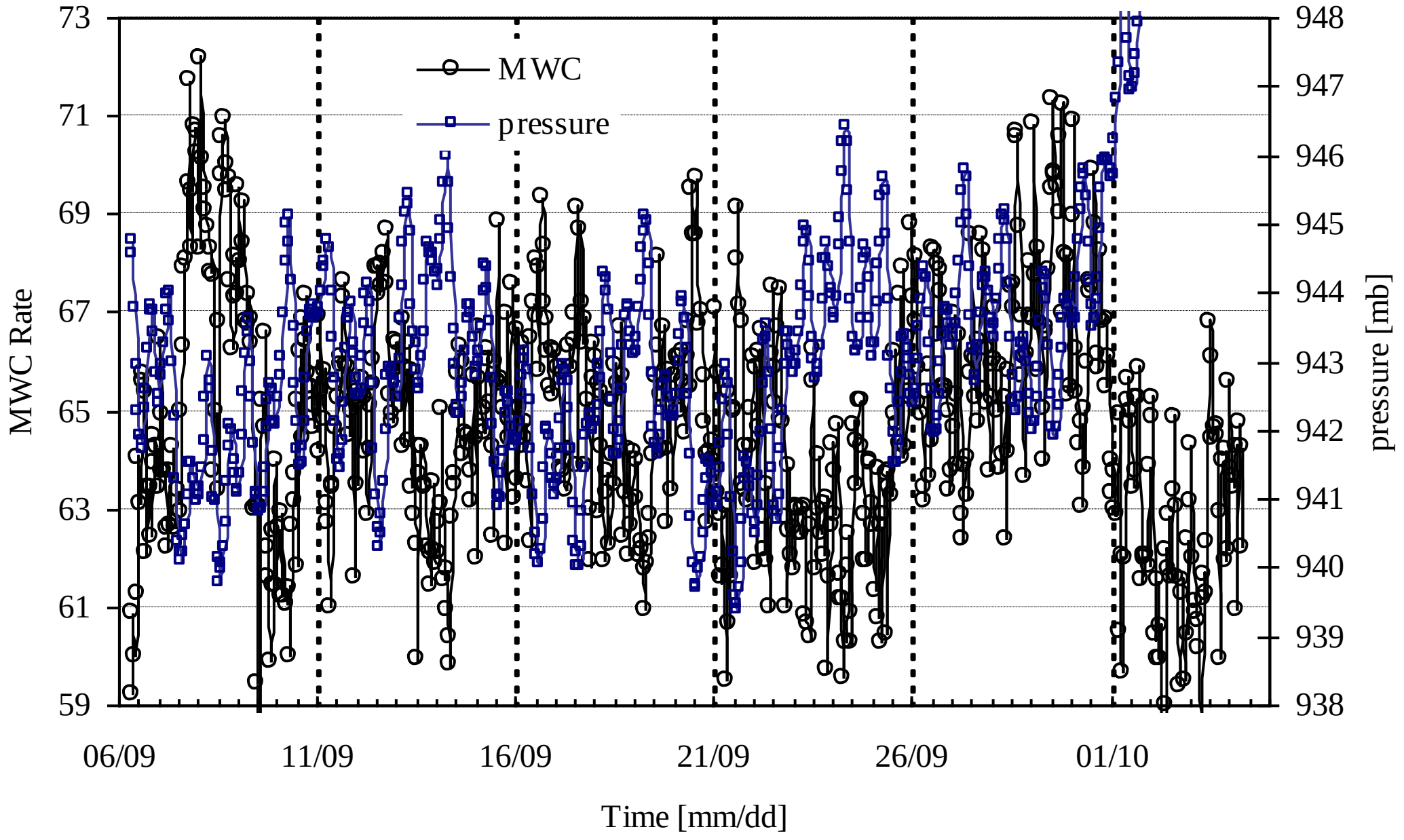

Figure 7: Time variations of the MWC muons and atmospheric pressure between 7th September and 3rd October 2015.

The pressure effect can be characterized by calculating the barometric coefficient. The changes of the detected secondary cosmic particles due to the atmospheric pressure can be expressed as follows [9]:

$$I = I_o exp^{[-\alpha(P-P0)]} \quad (1)$$

I is the cosmic ray rate at pressure P, and $I_0$ is the rate at the standard atmospheric pressure P0. The barometric coefficient, α, is usually determined experimentally by a simple linear regression between the muon intensities and the atmospheric pressure.

Figure (8) is a scatter plot comparing the atmospheric pressure and the muon rate during the whole study period. The slope of the linear fit (α) between these two variables yields the correction of d(Rate)/dp = (-0.765 ± 0.130) % / mb. For the same period of time (not shown here), a value of 0.181 ± 0.041 was obtained for the 1 $m^2$ detector [12]. This value does not differ much from that obtained previously [10 and 12]. It is noticeable that the α value from the MWC data is approximately five fold higher than that of the 1 $m^2$. There are several reasons for this difference of the response function of the MWC being different from that of the 1 $m^2$. Detailed investigations of this issue can be done by collecting data from the MWC for a longer period of time. For the purpose of the presented work, we corrected the recorded muons using the obtained $\alpha$ value and with aid of equation (1).

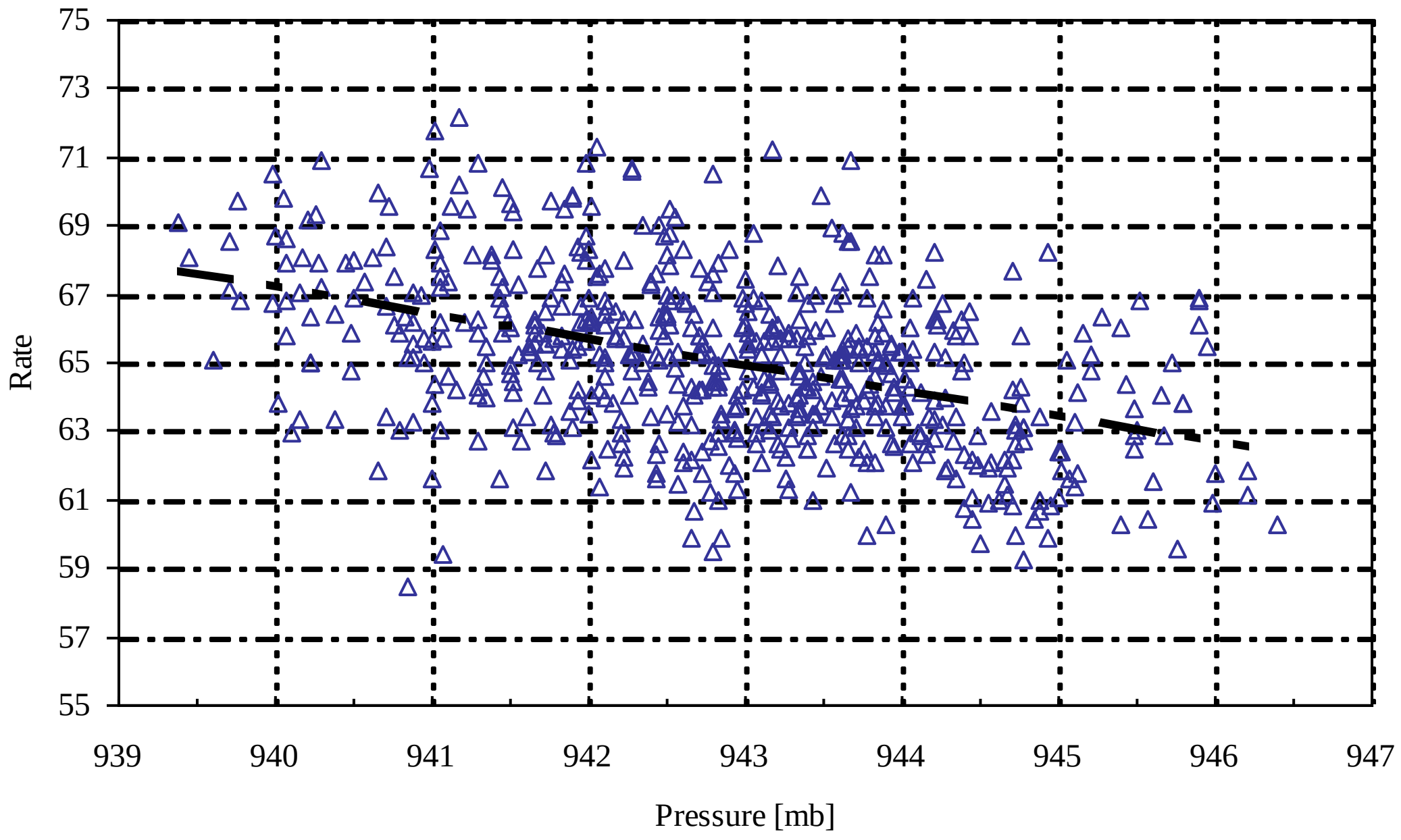

Figure 8: the muon count rate versus the atmospheric pressure.

Figure 9a compares the hourly pressure corrected and the uncorrected muons observed by the MWC detector. Figure 9b shows the daily averages of the pressure corrected muons for both the MWC telescope and the 1 $m^2$ scintillator. From both figures, it can be seen that this correction removes variations caused by changing atmospheric conditions.

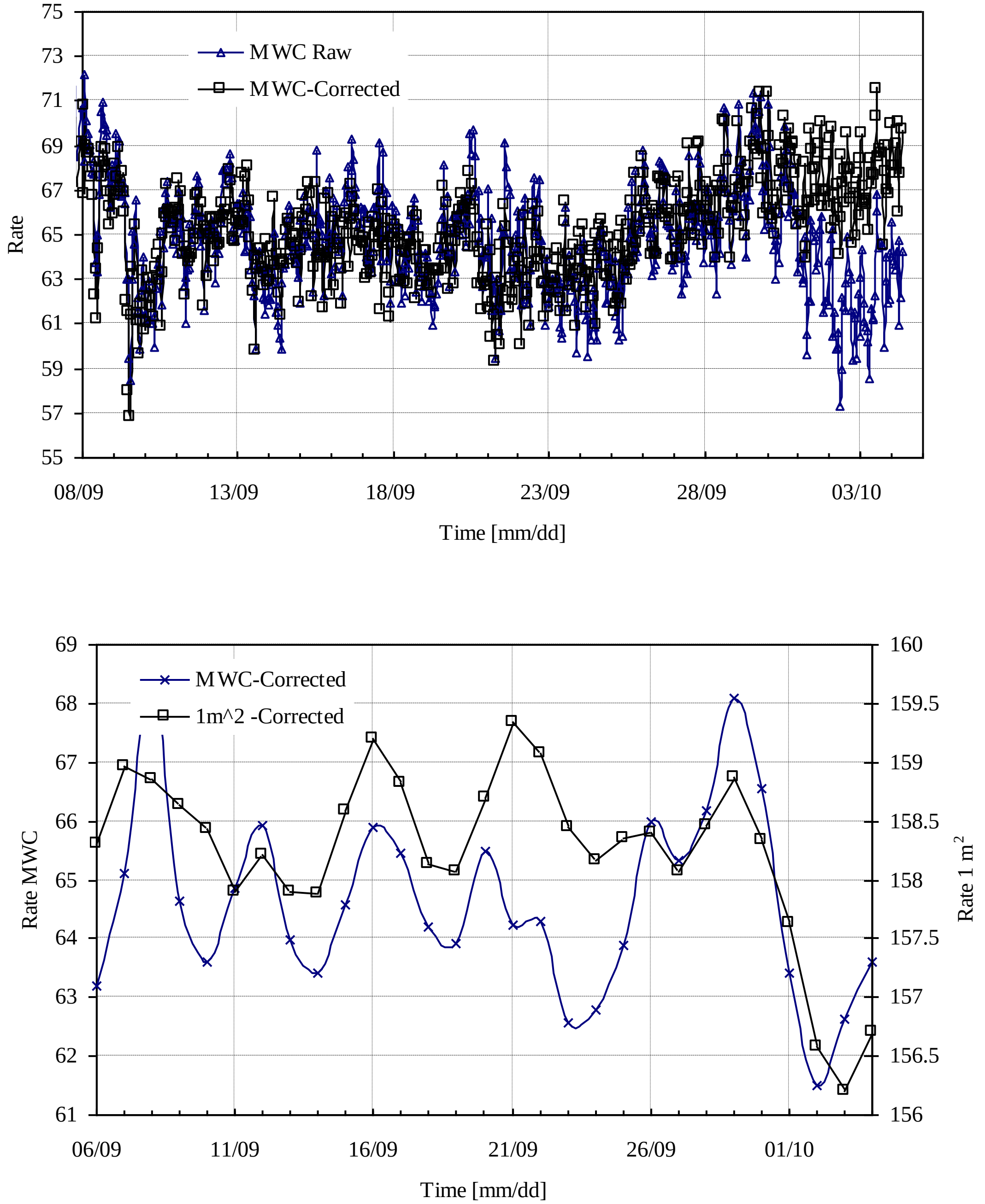

Figure 9: (a) the hourly variation of the raw muon count data and the pressure corrected count rates for a period of five days and (b) daily variations of the corrected muons from both detectors.

The effect of the atmospheric temperature is treated in a similar manner as the pressure effect [11-13]. The temperature coefficient can be calculated using Equation (2):

$$I = I_o exp^{[-\beta(T-T0)]} \qquad (2)$$

β is the temperature coefficient, and T and To are the measure and mean temperatures during the study period.

Similar regression analyses between muon rate and temperature were carried out to obtain the temperature coefficient. Figure (10) shows the time variations between the air temperature and pressure corrected muons detected by (a) the MWC telescope and (b) the 1 $m^2$ scintillator detector. It can be seen that the temperature variations slightly affect the cosmic ray muons, in comparison with the pressure effect. It is also noticeable that the muon rates increase as the temperature increases (positive correlations). The temperature coefficients from the MWC and scintillator detectors are d(Rate $_{MWC}$)/dT = (+0.0537 ± 0.0053) % / T and d(Rate $_{1\,m^2}$) /dT (+0.0837 ± 0.0086) % / T respectively.

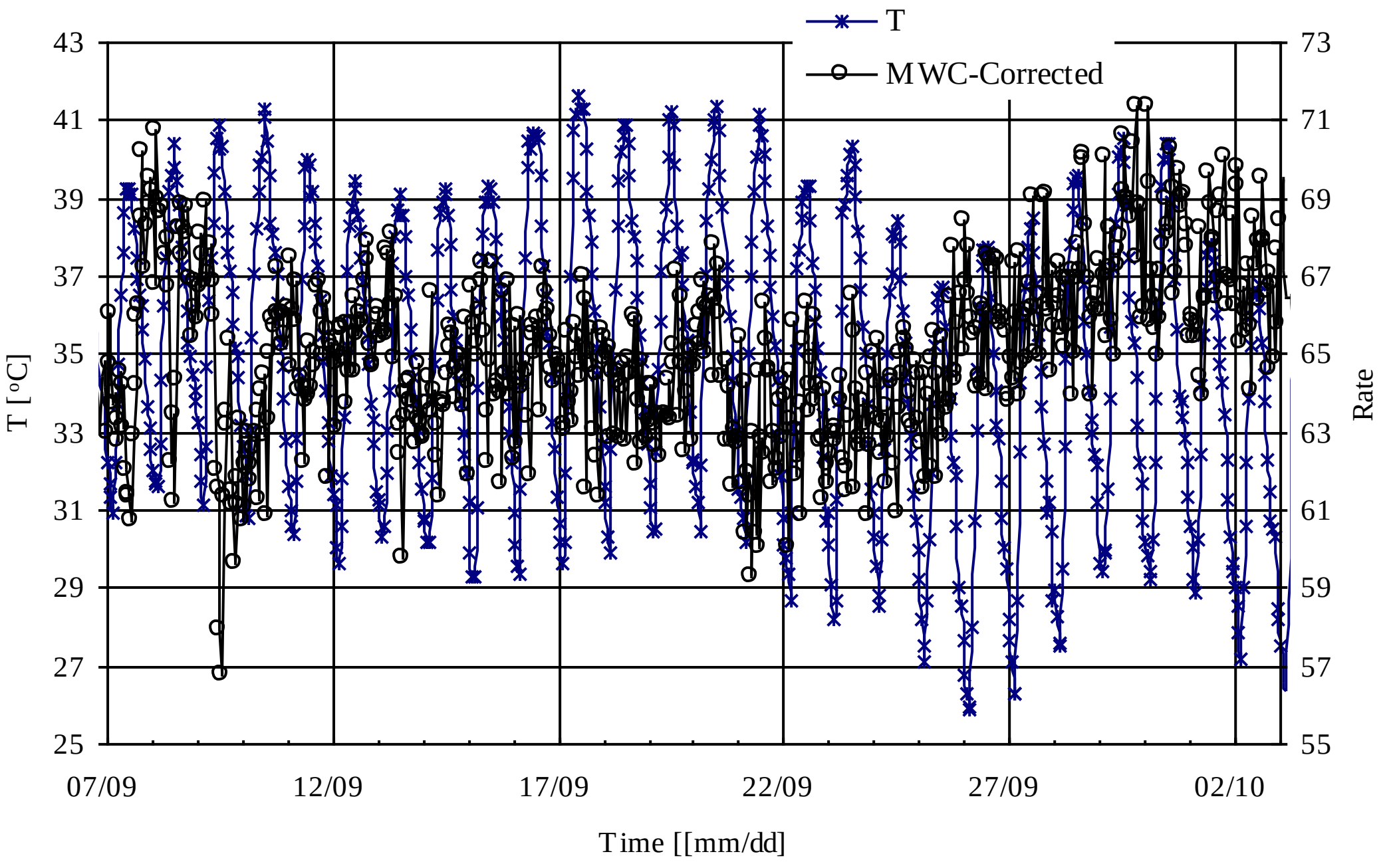

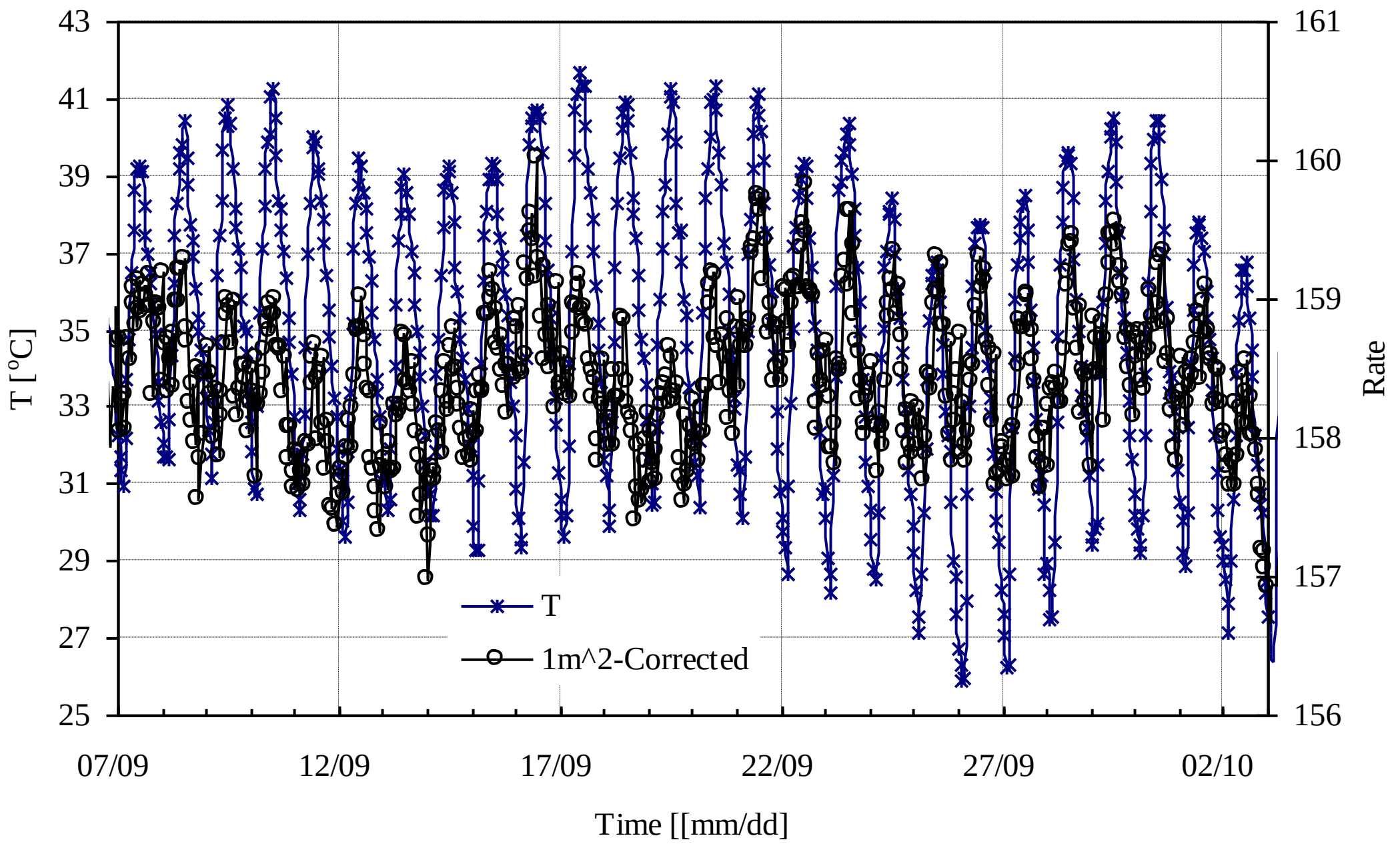

Figure 10: Time variations between the air temperature and pressure corrected muons detected by (a) the MWC telescope and (b) the 1 $m^2$ scintillator detector.

By running our developed detector for a longer period of time and maintain its performance, we will be able to study the atmospheric and heliospheric modulations of cosmic ray muons in more details [14].

## 4. Conclusions

A small MWC telescope (20×20 cm) was constructed and operated in Riyadh, Saudi Arabia to monitor the variations of cosmic ray muons. The telescope record high energy muons produced by primaries with energies higher than 10 GeV. The constructed telescope has effective detection efficiency, low power consumption and low cost. Although it has a small area, its ability to detect cosmic ray muons was comparable with the data obtained from a large area (1 $m^2$) scintillator detector running at the same site. Data from such a type of cosmic ray detector can be used to complement neutron monitor and muon detector observations at different places around the world with different energy ranges. In our future plan, we aim to build a larger size MWC detector and compare its performance and results with the developed small telescope and our 1 $m^2$ muon system, as well as with detectors from other places around the world.

## Acknowledgments

The authors would like to thank King Abdulaziz City for Science and Technology (KACST) for supporting this work.